# Improving Resources Management in Network Virtualization by Utilizing a Software-Based Network

Amir Javadpour[1,*]

[1]School of Computer Science and Technology, Guangzhou University, Guangzhou, China, 510006

*Correspondence to: a_javadpour@gzhu.edu.cn

*Abstract*— Network virtualization is a way to simultaneously run multiple heterogeneous architectures on a shared substrate. The main issue in network virtualization is mapping virtual networks to substrate network. How to manage substrate resources in mapping phase will have an effective role in improving the use of infrastructure resources. Using software-based networks in network virtualization which separates control logic from data as a new technology, has led to efficient resource management in this context. In this article a software-based network approach has been presented to network virtualization and manage infrastructure resources efficiently. It optimizes mapping function by dynamic resource management of infrastructure resource. We have added a module in the controller to manage the resources dynamically. An initial mapping will be done for arriving new requests based on number of successful requests and arriving time slots. They will not be finalized by writing the rules in the switches before arriving n requests. If some remapping during the n time window is needed, remapping can be done by the controller and the final results are sent to the switches to write the flow rules. The simulation has been done using NS2 simulator showed based on different evaluation criteria such as acceptance rate, average link utilization, cost and delay.

I. INTRODUCTION

Accelerating pace of computer network development has increased the need of using virtual networks to have better performance and lower costs in resource management. Network virtualization is a promising technology for the future of networks to add small networks together using programmable switches of Software Defined Networks (SDN) [1]–[5]. Whenever a new technology is introduced on the network, importance of its management is generally at its lowest level. As technology evolves and expands extensively, network management becomes a real need [2], [6]–[8]. Network virtualization is one of the most famous tools in networks which helps to manage resources and reduces general costs in the network[9]. The virtual network is a combination of active and passive network components (nodes and network links) on the substrate network. Virtual nodes connect to each other through virtual links in form of different topologies [2], [10], [11].

There are dynamic high flexible ways for network operators to manage or modify different type of networks [12]–[15]. The central controller in SDNs controls transforming equipment (such as switches etc.) and can be configured by an interface [10], [14], [16]. In the following three main parts of these networks is discussed. Application: it uses the decomposed levels of control and data to achieve specific goals such as security mechanisms or network measurement methods; Control plane: This level manages the sender's equipment with a controller to achieve the specific purpose of the target. Data plane: it supports the control level by using packets which are configured by controller.

Mapping problem can be divided into a node mapping and link mapping categories. Some approaches such as uncoordinated mapping maps node and link at different times. Node mapping might be performed earlier to provide some inputs arguments for link mapping. Uncoordinated mapping methods use a greedy algorithm to map the virtual node with highest resource request to the infrastructure node with the higher available resource. Lack of coordination between the node and link mapping might lead to an inefficient mapping. In other words, the selected infrastructure nodes might be physically dispersed and spread across the network infrastructure, which leads to more bandwidth consumption in infrastructure links. Considering the drawback of uncoordinated approaches, coordinated algorithms were proposed.

In some, nodes and links mapping has been done in two coordinate steps, in others both have been one in a same step. In a two-step coordinate mapping, the node mapping is performed considering the virtual network topology. In a one-step coordinated mapping, virtual links are mapped along with virtual nodes. It is possible to map links considering path division. In non-division methods, each virtual link is only mapped to one infrastructure path. In these cases, shortest path and k-shortest path algorithms will be used. The division of the path is further expressed in accordance with. In non-division methods by arriving the first request at t-1 and completing the mapping operation, it might be not enough resources to assign to the next request. But, considering the division methods, it is possible to resolve the need for a virtual link through several paths in infrastructure links. In summary, in division-based methods, the requested resources of virtual links will be covered by a proportion of resources through several paths in the infrastructure links. More requests will be accepted through this approach, but the packets may be received out of order. Mapping nodes and links are an important issue in network virtualization. It is important to consider the cost of using resources and have an optimized mapping solution [16]. Another important issue is resource management which can be considered dynamically or statically. In the dynamic management after the mapping, there is a possibility to perform a remapping for better use of network resources. While in a static management, resources devoted to requests remain unchanged until the end of the request's lives.

Receiving requests can either be online [17]–[20] or offline [16], [21]–[23]. In offline mode, all requests are available in the network before being mapped. In online mode, requests are received one by one and no information is available before any requests are received. An inefficient resource management can lead to misuse of resources which consequently will result in more rejected requests. On the other hand, an inappropriate resource management increases remapping requests. The purpose of this article is to reduce costs and delay by dynamic management of resources and considering requests online [24]–[27].

The following article will be presented as follows. In Section 2, a summary of related studies in the literature will be discussed. In Section 3, the issue will be addressed and the existing challenges will be expressed. Then, the network model and the approach on calculating the required parameters will be expressed. In Section 4, the proposed method is described. Section 5 contains the simulation procedure and evaluation of the results.

## II. RELATED WORK

Many studies have been done during last recent years to improve the quality of networks virtualization using software-based approaches and different strategies has presented to manage the resources. Most presented approaches assume that the network is static and all the requests have been delivered to the network offline. While some papers consider dynamic changes in the network [28], they did not consider node determination issue. Hoveydy et al [29] by proposing a flexible reconfiguration technique improved the performance for wide range of problems. The method proposed by Yo et al [18] considers the division of the substrate network for virtual link mapped onto multiple substrate paths. Also, the transfer route is intended to better use of substrate resources to increase the acceptance rate. In [19] Chowdhury et al, proposed two new mapping algorithms for virtual network requests. This approach have achieved a better correlation between the node and link mapping. Trivisono in [16] presented a static and offline mixed integer linear programming formulation for a central controller to provide optimal calculation of end-to-end virtual paths to the substrate internet provider which considers multiple requests simultaneously. As a result, the controller can improve end-to-end streaming management. Mahmet et al [22] have proposed a mapping approach in software based networks aimed to balance load on the substrate network and minimize delay for sending packets between controller and switch.

In [30] Feng et al proposed a method by combining link bandwidth allocations and flow tables for multiple control applications in a software based network. In addition, an integrated allocation model is built based on network resources in the software based networks to efficiently use link bandwidth with the minimum latency. Minjombi et al [24] instead of allocating a fixed amount of resources for the entire life of virtual networks(VN), dynamically and optimistically allocate resources to nodes and virtual links based on previous needs. The simulation results show that in the dynamic approach, the virtual network acceptance rate will be better than the static one. The proposed resource management system includes a source manager and a database which has expanded into software-based controller with better management. In this approach, improves the obtained results in [24] by synchronizing switch of resources and the links. It also reduces remapping by considering acceptance rate and the time window for the arrived requests. Ultimately the delay caused by remapping and costs will be decreased. Table I delivers a comprehensive comparison between several literature researches in different criteria.

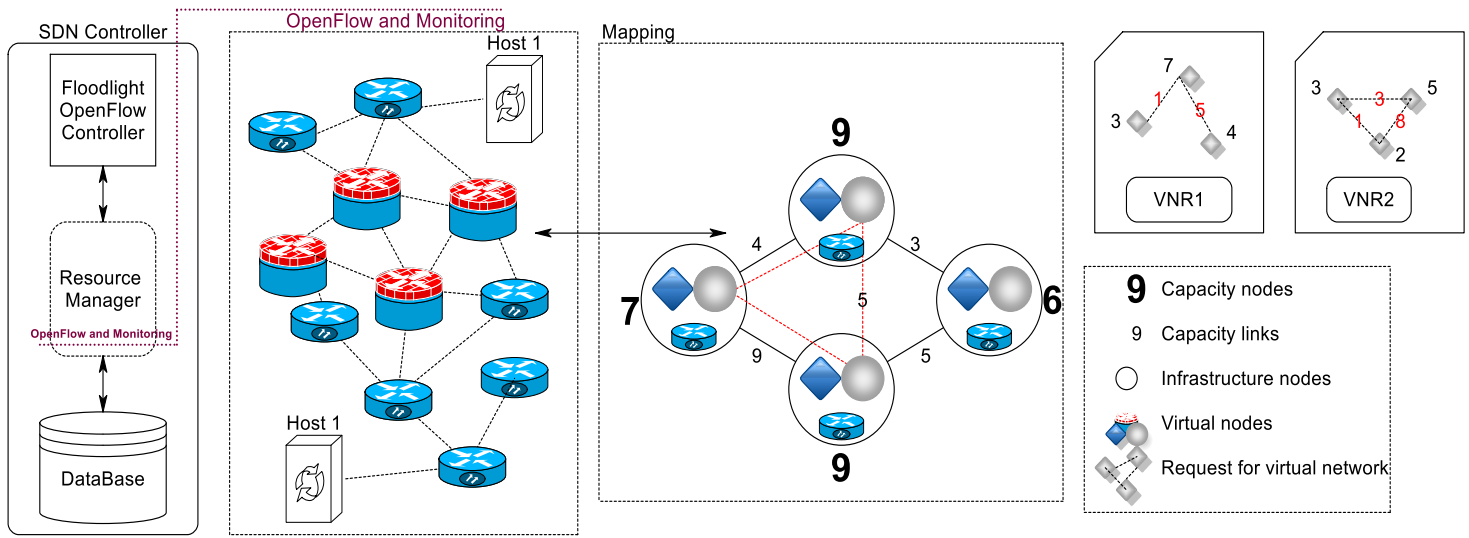

Fig. 1. Suggested method for network virtualization.

TABLE I. COMPARISON OF SOME PREVIOUS STUDIES AND RESEARCH STUDIES.

| Reference | Link/switch | Online/offline | Static/ dynamic | Goal |
|---|---|---|---|---|
| (Zhu & Ammar, 2006) | both | Offline | Relatively dynamic | Load balancing |
| (Houidi, Louati, Zeghlache, Papadimitriou, & Mathy, 2010) | link | Online | dynamic | Distributed error-tolerant algorithm for mapping |
| (Mijumbi, Serrat, & Gorricho, 2015) | both | Online | static | Self- Configuration  Self -Improvement |
| (Yu, Yi, Rexford, & Chiang, 2008) | both | Online | static | Increasing profits and reducing costs |
| (Chowdhury, Rahman, & Boutaba, 2012) | both | Online | static | Increasing acceptance and profit rates and reducing costs |
| (Trivisonno et al., 2013) | link | Offline | static | Optimal calculation of end-to-end virtual paths |
| (Demirci & Ammar, 2014) | both | Online | static | Fair allocation of link bandwidth and minimum total latency |
| (Feng, Bi, & Wang, 2014) | both | Online | dynamic | Increasing acceptance rates by guaranteeing the stability of service quality parameters |
| (Mijumbi R, Rubio-Loyola J, Bouten N, De Turck F, 2014) | both | Online | dynamic | Improving acceptance rates and reducing costs |
| Our method | both | Online | dynamic | Reducing costs and delays by ensuring the stability of the acceptance rate |

## III. RESEARCH STATEMENT OF PROBLEM

Network virtualization provides cost reducing solutions to the hardware needs of networks. Among various proposed approaches to allocate resources to network requests, the software-centric networks are an efficient way to manage the resources and requests. By separating the control logic from data, management of virtual networks and resources has been simplified. Although software-centric solutions are efficient, there are no adequate researches in the literature about them. In this section, a new algorithm using a software-based network has been presented which considers the capacity of links and time window for receiving the requests. In this section, we will first discuss embedding or mapping problem which is the most important topic in virtualization.

Different steps of storing data on the database using SDN controller and FloodLight is shown in figure 1. Links between the switches should be monitored to manage resources in SDN controller. Red switches are a part of SDN network which we are using as an example of OpenFlow and monitoring to illustrate how mapping problem will be applied on the network. Mapping section of figure 1 contains two virtual networks. A node in substrate network can host several virtual nodes and each substrate link can host multiple virtual links. One of the virtual links is mapped onto two substrate links and this suggests that a virtual resource can be a combination of several sources of substrate.

Several issues should be considered in the mapping: firstly, the candidate substrate resources in mapping must support the needs of virtual resources. For example, a virtual link MBits 1000 and MBits100 cannot be mapped onto substrate link. Secondly, the CPU request of a virtual node should also be less than the capacity of the node substrate that it wants to map. If more needs to be done, more substrate resources should be reserved but the consumption of substrate resources should be cost-effective and therefore optimal mapping. Sometimes it is necessary to carry out some mapping for optimal mapping and acceptance of additional requests or in other word, mapped requests to be remapped so that more resources are freed up and the resources cannot be split up.

### A. Network Virtualization

Efficient use of network infrastructure resources involves use of effective techniques to solve mapping problem. In general, mapping is an important issue in theory and in practice due to various topologies the virtual networks, resource restrictions regarding to network infrastructure, limited number of resources and dynamic online arriving requests. Although computational challenges of virtual network placement have led researchers to focus on heuristic strategies that limit the problem space in one or more dimensions, a linear programming approach has been used in this research. Virtual network mapping strategies and models are described below.

### B. Initial mapping of virtual network

There are several issues to consider such as available capacity, maximum route delays, geographic location, cost, etc. in order to set the initial mapping. In the following, the general model of mapping using infrastructure model of network, virtual network model and other parameters being used in the literature will be reviewed. An infrastructure network in general is a non-directed graph Gs = (Ns, Ls) which Ns is infrastructures nodes (ns) and Ls is infrastructures links (ls). For each node, $a_{n_s}$ Shows available capacity of node and cost (ns) is the cost of using each unit of source. $P_s$ is a set of infrastructure paths (without rings) between two nodes in Gs. $a_p$ is the minimum of bandwidth among all infrastructure paths. Virtual network model is also a non-directed graph Gv = (Nv, Lv). For each virtual node $n_v \in N_v$, $r_{n_v}$ represents the minimum requested capacity by virtual node. $r_{n_l}$ specifies the minimum bandwidth required for each virtual link. In Figure 2, an example of infrastructure network along with virtual network has been shown. The numbers inside the squares are the capacity of the existing nodes and the numbers next to the links expresses the available bandwidth of each link.

## C. Mapping model

Mapping model is defined by allocating proper infrastructure for arriving virtual network requests in a way to minimize the mapping costs. Assigned resources will be released upon completion of the tasks, and will be ready to respond to new requests. A mapping problem $\mathcal{M}$ can be divided into mapping node and mapping link sub-problems from Gs to Gv. The below conditions should be satisfied by a mapping problem. The infrastructure node which hosts a virtual node should has sufficient free capacity: Equation 1. Each virtual node can only be hosted by a one infrastructure node (Equation (2). Proper mapping $\mathcal{M}: L_v \to P_s$ or $l_v \mapsto \mathcal{M}(l_v)$ has to have a hosting path (Equation 3) and enough bandwidth for mapping (Equation 4). Consuming the infrastructure resources by allocated requests is defined as cost of network. Therefore, the mapping cost is equal to the total resources allocated to the virtual network and can be calculated through Equation 5. According to this relation, total cost is equal to the sum of all allocated capacities of the host infrastructure to the virtual nodes. Table III summarizes all used variables with the definition.

$$a_{\mathcal{M}(n_v)} \geq r_{n_v}, \forall n_v \in N_v \qquad \text{Equation (1)}$$

$$\mathcal{M}(n_v) = \mathcal{M}(m_v) \text{ iff } n_v = m_v, \forall n_v, m_v \in N_v \qquad \text{Equation (2)}$$

$$\mathcal{M}(l_v) = p. \exists p \in P_s \forall l_v \in L_v \qquad \text{Equation (3)}$$

$$a_p \geq r_{l_v} \qquad \text{Equation (4)}$$

$$\text{Cost}(G_v) = \sum_{n_v \in N_v} \text{cost}(\mathcal{M}(n_v)) * r_{n_v} + \sum_{l_v \in L_v} \sum_{l_s \in \mathcal{M}(l_v)} \text{cost}(l_s) * r_{l_v} \qquad \text{Equation (5)}$$

## D. Network model

Substrate networks are modeled with non-directional weighted graphs which are represented as G(S,L) while S represents the substrate switch and L represents the substrate link. Each substrate link G'(S',L') is a member of the set of l that connects the u and v substrate switches and has a $B_{uv}$ bandwidth. The capacity of each substrate switch like u ($u \in S$) is shown with $M_u$. Similarly, in the virtual network, G'(S',L') shows the virtual network topology which S' represents a virtual switch set and L' represents a virtual link set. Each virtual link $l'_{u'v'}$ ($l'_{u'v'} \in L'$) that connects the 'u' and 'v' substrate switches and has a bandwidth $B'_{u'v'}$. The capacity of each infrastructure switch such as u' ($u' \in S'$) is represented by $M'_{u'}$ and $P_{l u'v'}$ is the substrate path to which each virtual link is mapped.

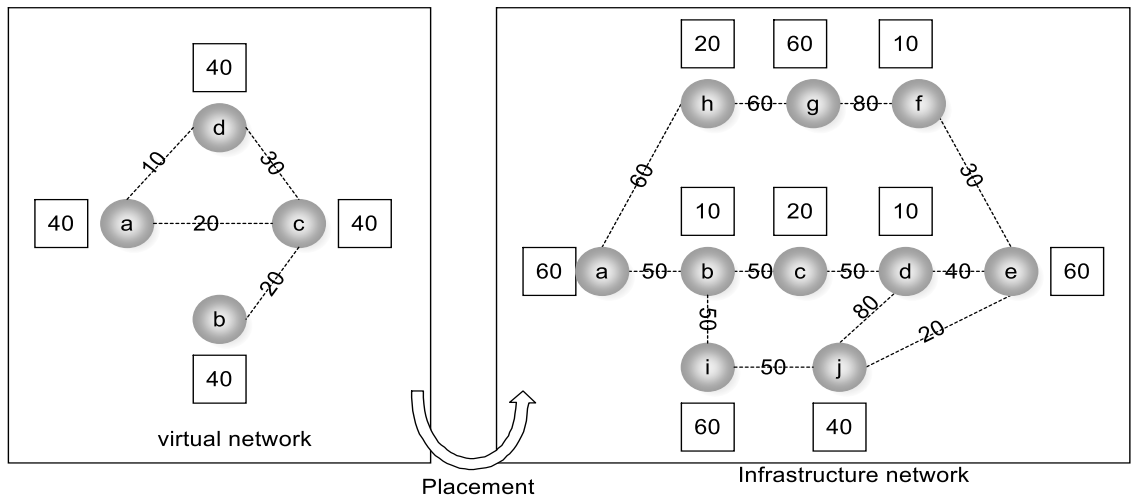

Fig. 2. Infrastructure and Resources Network and Link Infrastructure.

## E. Required parameters of the proposed method

In this article, the proposed method is used to calculate the weight of the virtual links by Minjombi et al [24], that in the following, they will be listed in Table II. For each virtual link $l_{u'v'}$ the parameter $R_{u'v'}$ represents the total amount of substrate resources used through the virtual link mapping. It is equal to the total amount of bandwidth and memory in the desired path of virtual links (Equation 6).

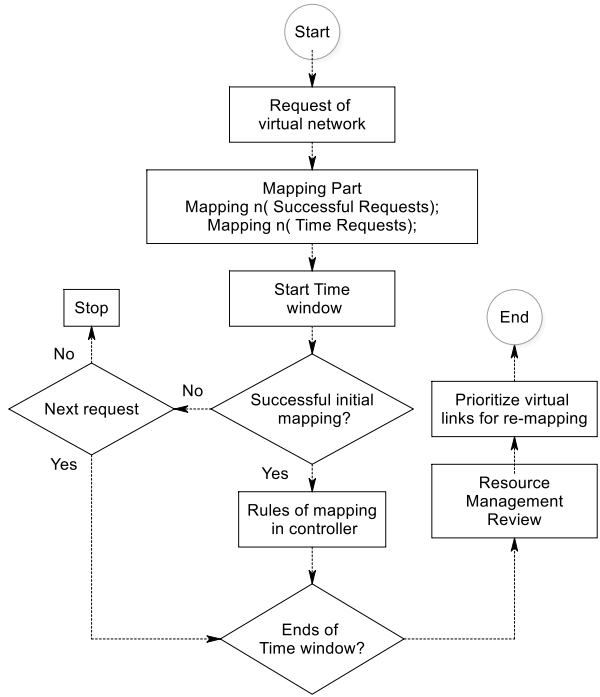

*The flowchart of Proposed Algorithm*

Fig. 3. Suggested method for network virtualization.

*Proposed Algorithm*

**Start;**
**Waiting for request Vn;**
Initial mapping is done on infrastructure;
The resources that is available on the network;
**If** enough resources are available successful;
   *Mapping for n(Requests);*
   *Mapping for n(Time);*
   **Check** *The amount of link resources and switches used by the substrate*
   **Check** *The amount of link sources and free and*
   **Check** *unchecked switches of substrate*
   **Check** *Weight assigned to any virtual link*
   **Then**
   Check Dynamic resource management;
   *A module is activated in the controller;*
   *A series of calculated by weight;*
   *Calculation of the parameters required for the proposed method*
   **Check unallocated resources substrate network**
   **Check link mapping;**
   **Then** *next Request.*
**Else**
   **Then** *next Request.*
**End.**

Fig. 4. The proposed algorithm for network virtualization.

$$R_{u'v'} = \sum_{l_{uv} \in P_{l_{u'v'}}} (B_{u'v'}) + \sum_{u \in P_{l_{u'v'}}} (M_{u'v'}) \qquad \text{Equation (6)}$$

The other important parameter is the amount of unallocated resources of the substrate network. For each virtual link, the parameter $A_{u'v'}$, which represents the total amount of link resources and unallocated switches in the infrastructure network, has been computed so that for future mappings, selected resources have more load balances. This parameter is derived from the Equation 7. By calculating the link resources and unallocated resources, it is possible to compute weight of virtual links using Equation 8. Then, by descending sorting out the calculated weight of virtual links, it would prioritize a link with more weight to reduce the burden imposed on the substrate network.

$$A_{u'v'} = \sum_{l_{uv} \in P_{l_{u'v'}}} (B_{uv} - B_{u'v'}) + \sum_{u \in P_{l_{u'v'}}} (M_{uv} - M_{u'v'}) \qquad \text{Equation (7)}$$

$$W_{u'v'} = R_{u'v'} - A_{u'v'} \qquad \text{Equation (8)}$$

## F. Proposed method

In the proposed method, the network virtualization process will be performed and controlled using a software-based network. The model contains a written module in controller for managing resources. This module contains written rules about managing arriving requests and available resources. The proposed algorithm first waits for arriving of n requests to map. If sufficient resources are available to respond to the request, a successful initial mapping will be done otherwise the request will be rejected and awaiting further requests. If sufficient resources are available to respond to the request, a successful initial mapping will be done otherwise the request will be rejected and awaiting further requests. Requests with successful initial mapping will situate in a queue to make the n initial mapping. By performing the nth early successful mapping, the management module is triggered to calculate mapping parameters such as link's weight. The main difference of the proposed method and Minjombi et al. [24] is adding a time window for successful early mappings. In addition, the flow rules are not written in the switches immediately, but they are waiting in the mapping queue and no rule will be written in any switch before reaching n successful mapping and considering the arrival time of requests. All rules of the n successful mapping are written together at the same time in the switches. Flowchart of the proposed method is shown in Figure 2. The weights obtained for each virtual link have been recalculated and arranged by changing the mode of the network substrate resources, such as mapping or remapping or retrieving resources after the expiration of a request. The presented algorithm of this method is presented in Figure 3.

TABLE II. PARAMETERS REQUIRED TO CALCULATE THE WEIGHT OF EACH VIRTUAL LINK.

| Parameters | Explanation |
|---|---|
| $R_{u'v'}$ | The amount of link resources and switches used by the substrate |
| $A_{u'v'}$ | The amount of link sources and free and unchecked switches of substrate |
| $W_{u'v'}$ | Weight assigned to any virtual link |

IV. EVALUATION

To simulate the proposed algorithm, we have used Network simulator Ns2 version 2.35[31]. We have extended a FloodLight controller to include SDN module. Our library SN is created in Mininet. HyGenICC module incorporated at the link roots in properties topology setup and the SDN-GCC framework (i.e, the controller foundation and smooth-layer). In addition, we have patched ns2 using the publicly available TCP-SDN patch. Links speed are of 1 Gb/s for sending stations, a bottleneck link of 1 Gb/s, low round trip time is 100 μs and the default Transmission Control Protocol Real Time (TCPRT) of 200 ms have been used through the simulation. FloodLight and Mininet run on different Ubuntu 14.04 virtual machines each with 1.0GB RAM. The substrate network topology is based on the GEANT network. Virtual network topologies are created using Brite with a uniformly distributed number of nodes between 3 and 10. The virtual to substrate mapping is performed using node mapping with a greedy approach and link mapping using a multi-commodity flow (MCF) [22] formulation (without path splitting), which is then solved using CPLEX. The memory and bandwidth capacities of substrate switches and links are uniformly distributed between 100 and 250 units respectively. We assume Poisson arrivals at an average rate of 1 per 5 time units. The average service time of the requests is 120 time units and assumed to follow a negative exponential distribution[14], [32]. Table III illustrates the considered parameters in the simulation. Node mapping was carried out by implementing a greedy approach. Multi-commodity flow without path splitting is used to map links. The memory capacity of the switches and the bandwidth of the links are also considered with a uniform distribution between 100 and 250 units. Number of switches is defined 14 by default. In this article, it is assumed that each current rule consumes 1 unit of the memory of switches. The number of virtual network requests in simulation is up to 1500 requests.

Obtained results has been compared with two other approaches. SSPSM approach [18] uses a time window to receive requests. It also uses path splitting and link transfer for mapping. The SDN-VN approach [24] has also used SDN networks to manage and control resources and requests. While the proposed method in this article beside using SDN networks, benefits from time window for more effective initial mapping. It helps the algorithm to have less remapping. Figure 5 compares the acceptance rate of two literature researches and our proposed method based on n successful requests (Part A) and time (Part B). As it is illustrated the acceptance rate of our approach has significantly increased compared to the SSPSM slightly compared to SDN-VN. This is because of the possibility of remapping in the SDN-VN and our proposed method. It is because of considering n successful online requests instead of considering total number of arriving ones. This is related to the proposed time window in the proposed method. As the proposed method is based on online mapping, we decided to consider n successful mapping to overcome bad mappings. It should be noted that different runs of algorithm would have different results as the requests are considered online and the algorithm does not know about the upcoming requests.

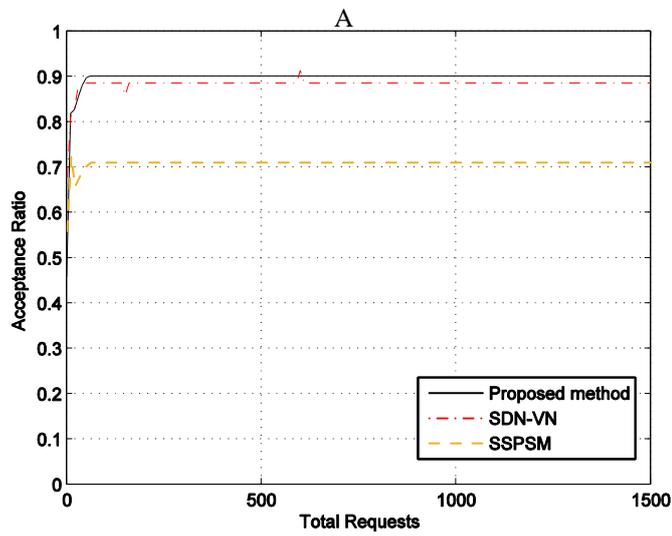 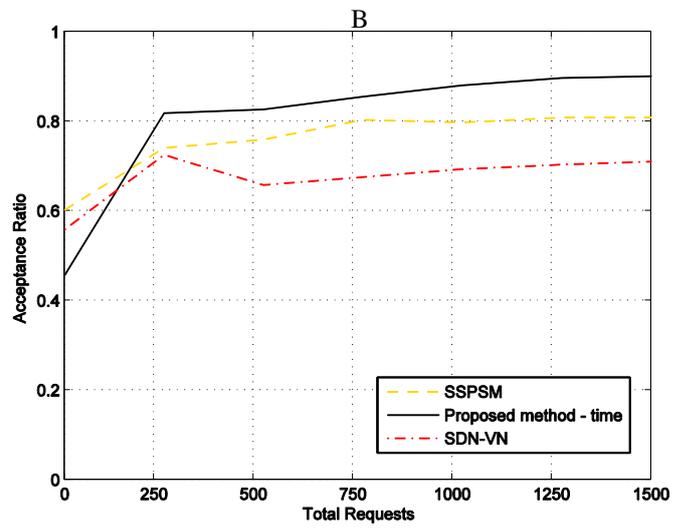

Fig. 5. The evaluation results of the acceptance rate; A is mapping based on n successful requests and B is based on time.

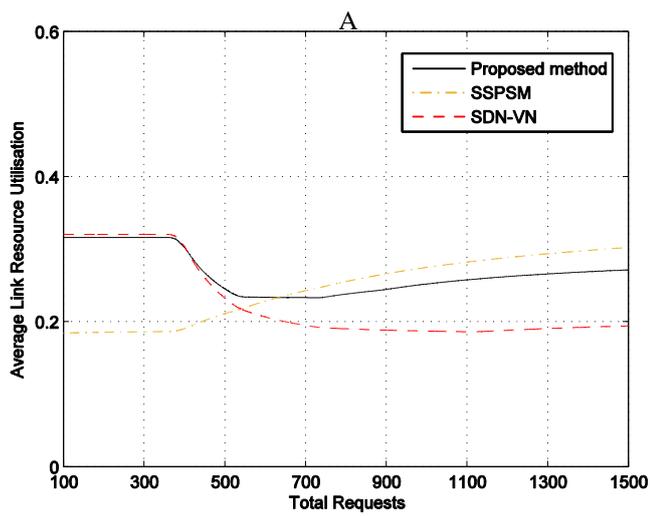 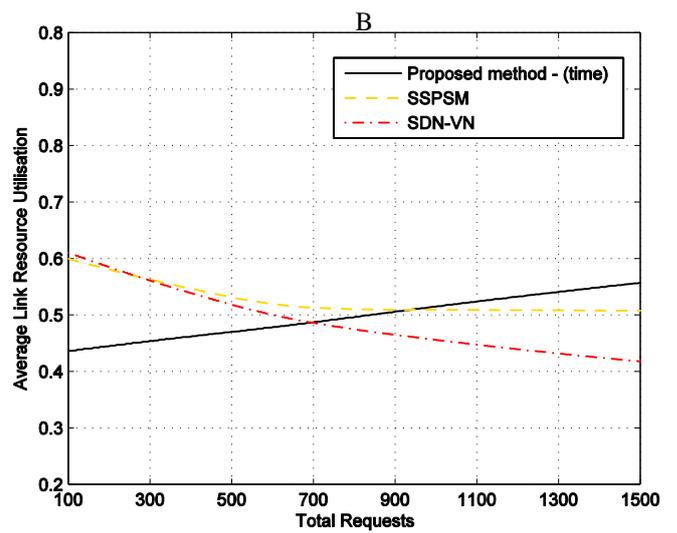

Fig. 6. Evaluation results of the average link resource utilization; A is mapping based on n successful requests and B is based on time.

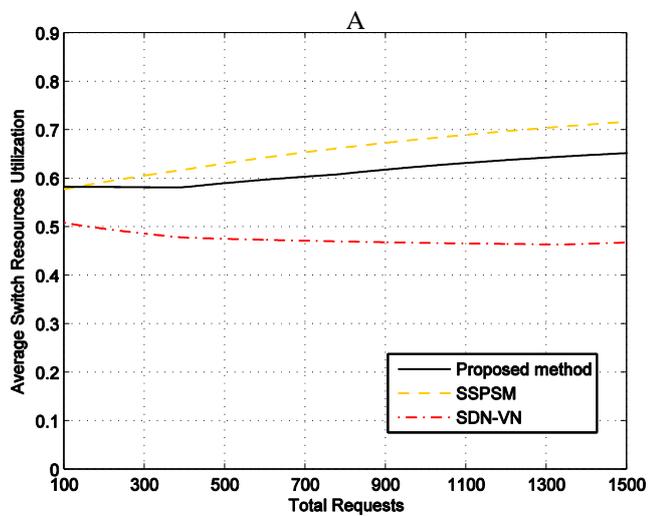 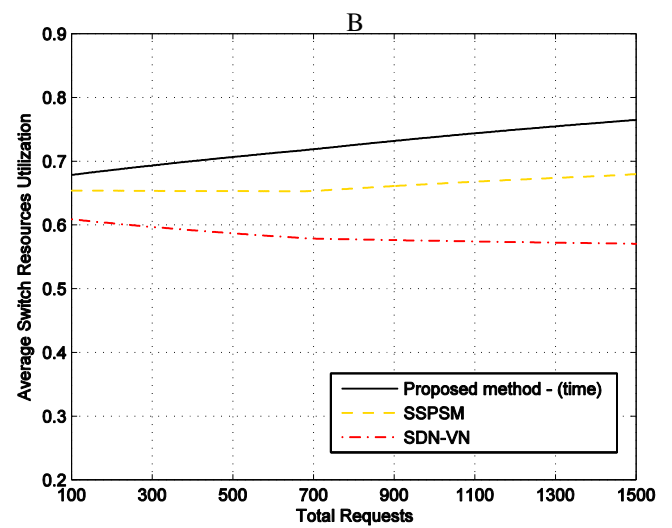

Fig. 7. Evaluation results of average switches resource utilization; A is mapping based on n successful requests and B is based on time.

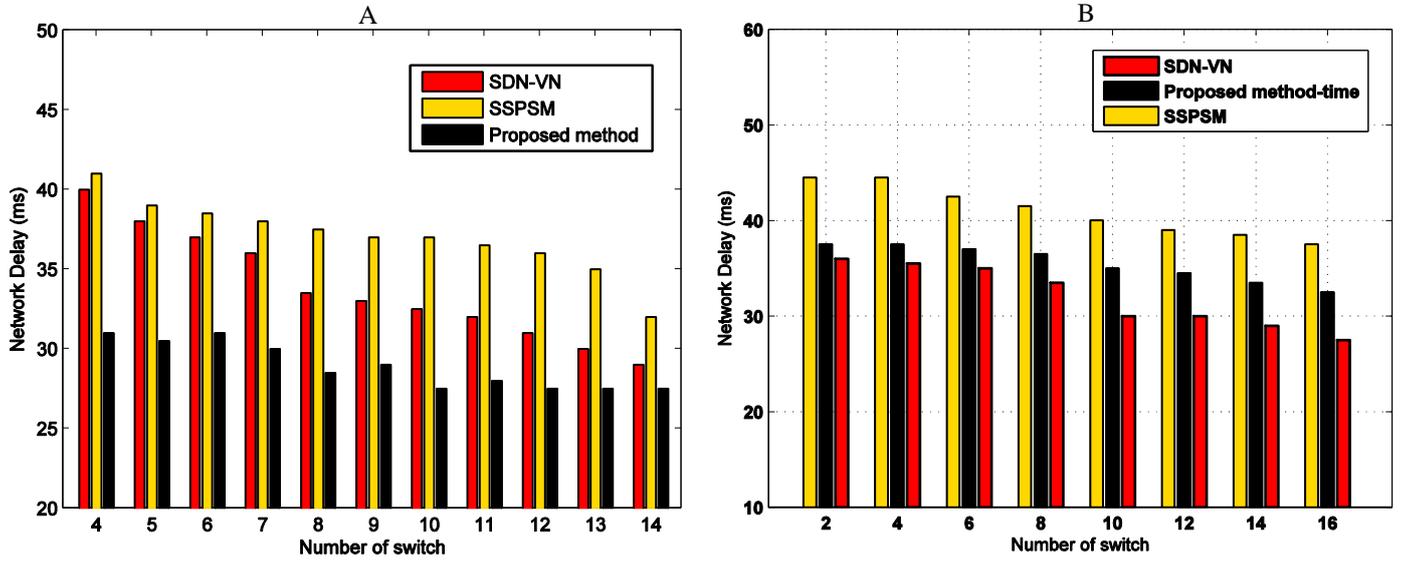

Fig. 8. Results of network latency evaluation; A is mapping based on n successful requests and B is based on time.

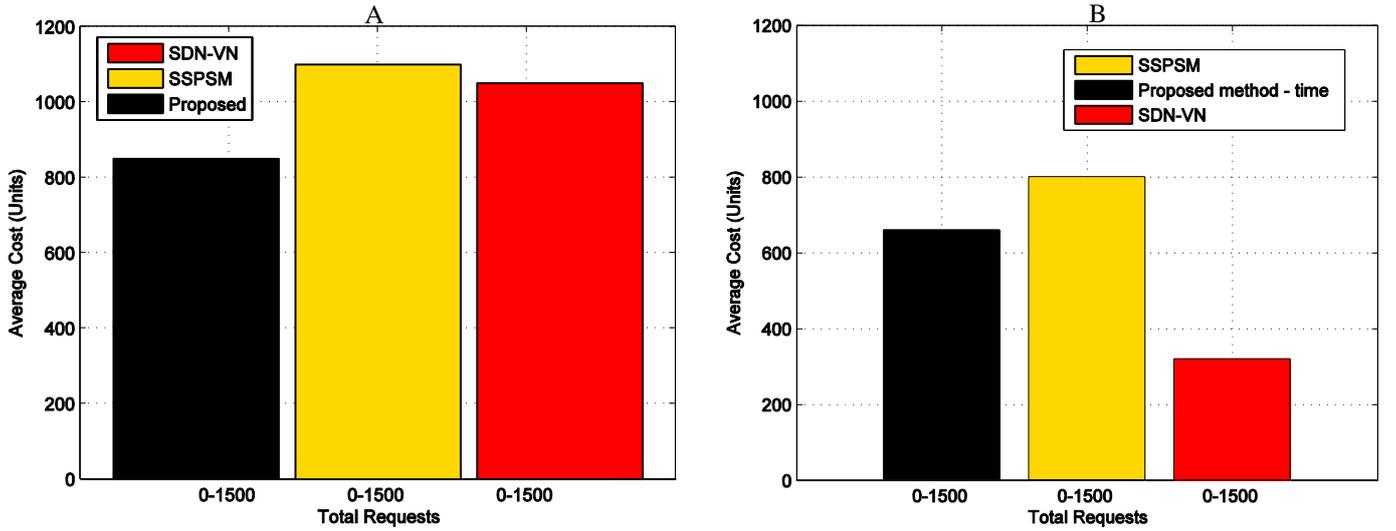

Fig. 9. The average cost in the network after virtualization; A is mapping based on n successful requests and B is based on time.

TABLE III. SIMULATION PARAMETERS.

| The parameters used in the NS2 simulator | Values in the simulator | The parameters | Values |
|---|---|---|---|
| loaded server | LS | Infrastructure network Topology | Gs |
| loaded Server selection | LSS | Infrastructure set of nodes | Ns |
| Periodic Synchronizations | PS | Infrastructure set of Links | Ls |
| Network Representation Mapping | NRMM | Virtual network topology | Gv |
| Management | MC | Virtual network set of nodes | Nv |
| Multiple controller | NRMR | Virtual network set of links | Lv |
| Network Representation Mapping Reserved | NRSM_R | Available capacity of infrastructure node $n_s$ | $a_{n_s}$ |
| Network Representation Sub Mapping RTS | NRSM_C | Cost of infrastructure node $n_s$ | cost (ns) |
| Network Representation Sub Mapping CTS | NRSM_A | Capacity of infrastructure link $l_s$ | $a_{l_s}$ |
| Network Representation Sub Mapping ACK | NRSM_D | infrastructure paths in Gs | $P_s$ |
| Network Representation Sub Mapping Data | NRSM_M | Available bandwidth in $P_s$ | $a_p$ |
| | | requested capacity of virtual node | $r_{n_v}$ |
| | | Requested bandwidth of virtual link | $r_{l_v}$ |
| | | Mapping function | $\mathcal{M}$ |

Figures 6 and 7 illustrate the average utilization of link and switches resources are compared for three approaches. These parameters indicate the average proportion of using resources. Although the results of our approach in figures 6,7 are not optimized over two references based on n request mapping, its performance is getting better by increasing the number of requests for time-based mapping while increasing number of requests in the other two approaches leads to a lower link utilization due to the static management in SSPSM. The management in the proposed method and SDV-VN are dynamic.

Figure 8 contains the comparison results for network latency. It has been demonstrated that network latency has dropped by increasing number of switches. Considering time window in the proposed method, the reduction rate has been reduced. In Figure 9, the mapping cost is compared. Mapping cost has been derived from Equation (5) which represents the total amount of resources in virtual request mapping. It is seen that the cost parameter in the proposed method of this article has been reduced compared to the above two approaches. Considering n successful requests causes a reduction in computation costs and it affects the whole cost of mapping.

V. CONCLUSION

Software Defined Networking (SDN) is a way to solve some networks and problems such as resource management, security, managing complexity, multi-casting, load balancing, and energy efficiency. SDN is an architectural platform which separates the control and data planes of a networking device and makes it feasible to control, monitor, and manage a network from a centralized node (the SDN controller). What is outlined in this article is to provide a dynamic management approach using SDN to better mapping the requests of virtual networks to the substrate network. The proposed method by using a resource management module in the controller (which is actually a mode-operation machine), provides the possibility of remapping for arriving requests. Also, by using the controller, only the modified rules in the corresponding switches will be changed. In other words, it is not necessary for the proposed approach to continuously update of all network switches. We also considered time window which leads to cost and network latency reduction. By increasing the number of mappings, the proposed method is more successful to reduce costs compared to other papers in the literature.

ACKNOWLEDGMENTS

This work is supported in part by the National Natural Science Foundation of China under Grants 61632009 & 61472451, in part by the Guangdong Provincial Natural Science Foundation under Grant 2017A030308006 and Hgh-Level Talents Program of Higher Education in Guangdong Province under Grant 2016ZJ01.

REFERENCES

[1] A. Blenk, A. Basta, M. Reisslein, and W. Kellerer, "Survey on Network Virtualization Hypervisors for Software Defined Networking," pp. 1–32, 2015.
[2] D. Kreutz and F. Ramos, "Software-Defined Networking: A Comprehensive Survey," *arXiv Prepr. arXiv ...*, pp. 1–61, 2014.
[3] T. Anderson, L. Peterson, S. Shenker, and J. Turner, "Overcoming the Internet impasse through virtualization," *Computer (Long. Beach. Calif).*, vol. 38, no. 4, pp. 34–41, 2005.
[4] A. Bavier, N. Feamster, M. Huang, L. Peterson, and J. Rexford, "In VINI Veritas: Realistic and Controlled Network Experimentation," in *Proceedings of the 2006 Conference on Applications, Technologies, Architectures, and Protocols for Computer Communications*, 2006, pp. 3–14.
[5] T. Hafeez, N. Ahmed, B. Ahmed, and A. W. Malik, "Detection and Mitigation of Congestion in SDN Enabled Data Center Networks: A Survey," *IEEE Access*, vol. 6, pp. 1730–1740, 2018.
[6] J. A. Wickboldt, W. P. De Jesus, P. H. Isolani, C. B. Both, J. Rochol, and L. Z. Granville, "Software-defined networking: management requirements and challenges," *IEEE Commun. Mag.*, vol. 53, no. 1, pp. 278–285, 2015.
[7] R. Kanagevlu and K. M. M. Aung, "SDN controlled local re-routing to reduce congestion in cloud data center," *Proc. - 2015 Int. Conf. Cloud Comput. Res. Innov. ICCCRI 2015*, pp. 80–88, 2016.
[8] Amir Javadpour and A. reza Mohammadi, "Improving Brain Magnetic Resonance Image (MRI) Segmentation via a Novel Algorithm based on Genetic and Regional Growth," *J. Biomed. Phys. Eng.*, vol. 6, no. 2, pp. 95–108, Jun. 2016.
[9] M. Karakus and A. Durresi, "Quality of Service (QoS) in Software Defined Networking (SDN): A survey," *J. Netw. Comput. Appl.*, vol. 80, pp. 200–218, 2017.
[10] D. Mithbavkar, H. Joshi, H. Kotak, D. Gajjar, and L. Perigo, "Round Robin Load Balancer using Software Defined Networking ( SDN )," *Capstone Team Res. Proj.*, 2016.
[11] R. Khondoker, a Zaalouk, R. Marx, and K. Bayarou, "Feature-based Comparison of Software Defined Networking (SDN) Controllers," *Int. Conf. Comput. Softw. Appl. 2014 Proc.*, 2014.
[12] M. Qilin and S. Weikang, "A Load Balancing Method Based on SDN," *Proc. - 2015 7th Int. Conf. Meas. Technol. Mechatronics Autom. ICMTMA 2015*, pp. 18–21, 2015.
[13] M. F. Bari, A. R. Roy, S. R. Chowdhury, Q. Zhang, M. F. Zhani, R. Ahmed, and R. Boutaba, "Dynamic Controller Provisioning in Software Defined Networks," *Network and Service Management (CNSM), 2013 9th International Conference on*. pp. 18–25, 2013.
[14] R. Mijumbi, J. Serrat, J. Rubio-Loyola, N. Bouten, F. D. Turck, and S. Latré, "Dynamic resource management in SDN-based virtualized networks," in *10th International Conference on Network and Service Management (CNSM) and Workshop*, 2014, pp. 412–417.

[15] S. Rezaei, H. Radmanesh, P. Alavizadeh, H. Nikoofar, and F. Lahouti, "Automatic fault detection and diagnosis in cellular networks using operations support systems data," in *NOMS 2016 - 2016 IEEE/IFIP Network Operations and Management Symposium*, 2016, pp. 468–473.

[16] R. Trivisonno, I. Vaishnavi, R. Guerzoni, Z. Despotovic, A. Hecker, S. Beker, and D. Soldani, "Virtual Links Mapping in Future SDN-Enabled Networks," in *2013 IEEE SDN for Future Networks and Services (SDN4FNS)*, 2013, pp. 1–5.

[17] R. Mijumbi, J. Serrat, and J.-L. Gorricho, "Autonomic Resource Management in Virtual Networks," *CoRR*, vol. abs/1503.0, 2015.

[18] M. Yu, Y. Yi, J. Rexford, and M. Chiang, "Rethinking Virtual Network Embedding: Substrate Support for Path Splitting and Migration," *SIGCOMM Comput. Commun. Rev.*, vol. 38, no. 2, pp. 17–29, 2008.

[19] M. Chowdhury, M. R. Rahman, and R. Boutaba, "ViNEYard: Virtual Network Embedding Algorithms With Coordinated Node and Link Mapping," *IEEE/ACM Trans. Netw.*, vol. 20, no. 1, pp. 206–219, 2012.

[20] M. Z. A. Bhuiyan, J. Wu, G. Wang, T. Wang, and M. M. Hassan, "e-Sampling: Event-Sensitive Autonomous Adaptive Sensing and Low-Cost Monitoring in Networked Sensing Systems," *ACM Trans. Auton. Adapt. Syst.*, vol. 12, no. 1, pp. 1:1–1:29, 2017.

[21] Y. Zhu and M. Ammar, "Algorithms for Assigning Substrate Network Resources to Virtual Network Components," in *Proceedings IEEE INFOCOM 2006. 25TH IEEE International Conference on Computer Communications*, 2006, pp. 1–12.

[22] M. Demirci and M. Ammar, "Design and analysis of techniques for mapping virtual networks to software-defined network substrates," *Comput. Commun.*, vol. 45, no. Supplement C, pp. 1–10, 2014.

[23] K. Qin, C. Huang, N. Ganesan, K. Liu, and X. Chen, "Minimum cost multi-path parallel transmission with delay constraint by extending openflow," *China Commun.*, vol. 15, no. 3, pp. 15–26, 2018.

[24] L. S. Mijumbi R, Rubio-Loyola J, Bouten N, De Turck F, "Dynamic resource management in SDN-based virtualized networks," in *10th International Conference on Network and Service Management (CNSM) and Workshop*, 2014, pp. 412–417.

[25] A. Javadpour, S. Kazemi Abharian, and G. Wang, "Feature Selection and Intrusion Detection in Cloud Environment Based on Machine Learning Algorithms," *2017 IEEE Int. Symp. Parallel Distrib. Process. with Appl. 2017 IEEE Int. Conf. Ubiquitous Comput. Commun.*, pp. 1417–1421, 2017.

[26] A. Javadpour, H. Memarzadeh-Tehran, and F. Saghafi, "A temperature monitoring system incorporating an array of precision wireless thermometers," in *2015 International Conference on Smart Sensors and Application (ICSSA)*, 2015, pp. 155–160.

[27] A. Javadpour and H. Memarzadeh-Tehran, "A wearable medical sensor for provisional healthcare," *Physics and Technology of Sensors (ISPTS), 2015 2nd International Symposium on*. pp. 293–296, 2015.

[28] I. Houidi, W. Louati, D. Zeghlache, P. Papadimitriou, and L. Mathy, "Adaptive Virtual Network Provisioning," in *Proceedings of the Second ACM SIGCOMM Workshop on Virtualized Infrastructure Systems and Architectures*, 2010, pp. 41–48.

[29] a Haider, R. Potter, and a Nakao, "Challenges in resource allocation in network virtualization," *20th ITC Spec. Semin.*, no. May, 2009.

[30] T. Feng, J. Bi, and K. Wang, "Joint allocation and scheduling of network resource for multiple control applications in SDN," in *2014 IEEE Network Operations and Management Symposium (NOMS)*, 2014, pp. 1–7.

[31] T. Issariyakul and E. Hossain, "Introduction to network simulator NS2," *Introd. to Netw. Simulator NS2*, vol. 9781461414, no. https://www.springer.com/gp/book/9781461414056, pp. 1–510, 2012.

[32] A. M. Abdelmoniem and B. Bensaou, "Efficient Switch-Assisted Congestion Control for Data Centers: an Implementation and Evaluation," *Proc. IPCCC*, 2016.